\magnification=1200
\parskip 10pt plus 5pt
\parindent 14pt
\baselineskip=18pt
\input mssymb
\pageno=0
\footline={\ifnum \pageno <1 \else \hss \folio \hss\fi}
\line{\hfil{hep-th/9503231~~~~~~~~~~~~~}}
\line{\hfil{(Updated version)~~~~~~~~~~}}
\line{\hfil{July, 1995~~~~~~~~~~~~~~~~~~~~~}}
\vskip .75in
\centerline{\bf MODULAR INVARIANCE AND THE FINITENESS OF SUPERSTRING THEORY}
\vskip  .9in
\centerline{\bf SIMON DAVIS}
\vskip .4in
\centerline{Department of Applied Mathematics and Theoretical Physics}
\vskip 1pt
\centerline{University of Cambridge}
\vskip 1pt
\centerline{Silver Street, Cambridge CB3 9EW}
\vskip .5in
\noindent{\bf ABSTRACT}.  The genus-dependence of multi-loop superstring
amplitudes is bounded at large orders in perturbation theory using the
super-Schottky group parametrization of supermoduli space.
Partial estimates of supermoduli space integrals suggest an
exponential dependence on the genus when the integration region is
restricted to a single fundamental domain of the super-modular group in the
super-Schottky parameter space.  Bounds for N-point superstring scattering
amplitudes are obtained for arbitrary N and are shown to be consistent with
exact results recently obtained for special type II string amplitudes for
orbifold or Calabi-Yau compactifications.  It is suggested that the generic
estimates, which imply the validity of superstring perturbation theory in the
weak-coupling limit, might be used to determine scattering amplitudes at
strong coupling because of the S-duality of type II and heterotic string
theories.

\vfill
\eject

The combination of supersymmetry and the smooth geometry of the interaction
region in superstring theory has been shown to lead to finite scattering
amplitudes at each order in perturbation theory and vanishing multiloop amplitudes when there
are fewer than four external massless states.  These results may be contrasted
with
the divergent amplitudes and even the genus-dependence of regularized
amplitudes in bosonic string theory.  In particular, a lower bound for the
regularized closed bosonic string path integral was found to increase at a
factorial rate with respect to the genus.  This bound was based on an argument
typically used in field theory, the counting of non-isomorphic trivalent
graphs,
and appeared to be independent of the content of the string theory.  It has
been suggested that this counting of diagrams, and therefore the growth, could
persist in superstring theory [1].  However, the lower bound of [2] was
derived from an integral over Teichmuller space rather than moduli space
and does not suggest a connection between supersymmetry and large-order
string perturbation theory.  Similarly,  a counting of cells
in a triangulation of punctured moduli space yields a factorial growth with
respect to the genus [3], but this estimate will be reduced by the introduction
of a cut-off in moduli space, excluding those Riemann surfaces with closed
geodesics
having lengths less than some genus-independent bound.

An analysis of the regularized bosonic string vacuum amplitude in the Schottky
\hfil\break
parametrization has revealed
that the restriction to a single copy of the fundamental region of the modular
group reduces the regularized path integral by a genus-dependent factor [4].
Although the regularized integral increases rapidly with respect to the genus,
it does not grow at a factorial rate, thus leaving open the possibility of
a qualitatively different result for superstrings.  Moreover, a connection
is established between the genus-dependence of the limits for the Schottky
group parameters and the large-order behaviour of the amplitudes, suggesting
that the infrared and large-order divergences, which occur in the
infinite-genus
limit for the bosonic string, might be simultaneously eliminated in the
superstring path integral. The techniques of [4]
shall be adapted to superstring theory in this paper.  It will be demonstrated
that, based on a partial estimate of the bounded superstring amplitudes,
the maximal dependence on the genus may be exponential, revealing a higher
degree of finiteness in the large-order limit.

Superstring scattering amplitudes are given in general by supermoduli
space integrals
[5]
$$\eqalign{\langle V_1(k_1)~...~V_N(k_N) \rangle_g~=~\int_{s
{\cal M}_g}~d\mu_{sWP}
\left({{8 \pi sdet^\prime {\hat{\square_0}}}\over {sdet{\langle~{\hat
{\Psi_\alpha}}\vert {\hat{\Psi_\beta}}\rangle}}} \right)^{-5}
& (sdet
{{\hat{\cal P}}_1^{\dag}}{{\hat {\cal P}}_1})^{1\over 2}
\cr
&\langle\langle V_1(k_1)~...~V_N(k_N) \rangle\rangle_{\hat E}
\cr}
\eqno(1)$$
where $d\mu_{sWP}$ is the super Weil-Petersson measure, $\Psi_\alpha \in
Ker \square_0$, $\langle\langle ~\rangle\rangle$
represents the evaluation of the path integral over the scalar position
superfields $X^\mu$, and the integral is restricted to a $(3g-3 \vert 2g-2)$
complex-dimensional slice of super-Teichmuller space parametrized by the 
supergeometries $\{{\hat E}_M^A\}$.  In terms of superghosts,
$$\eqalign{\langle V_1(k_1)~...~V_N(k_N)\rangle_g~=~\int_{s{\cal M}_g} d^2 m_K
\int D(XBC) &\langle\langle V_1(k_1)~...~V_N(k_N) \rangle \rangle
{}~\prod_b \vert \delta(\langle \mu_b\vert B \rangle)\vert^2
\cr
&~\prod_k \vert \langle \mu_k\vert B \rangle\vert^2 e^{-I}
\cr}
\eqno(2)$$
where B and C are anti-ghost and ghost superfields of U(1) weight ${3\over 2}$
and $-1$ respectively, $\mu_K,~ K=(k,b)$, are super Beltrami differentials, and
$I=I_m+I_{sgh}$  is the sum of the matter and superghost actions [5].  A
formula similar to that given in equation (2)  for the scattering
amplitude of N external massless states appears in the twistor-string 
formalism [6].

The Schottky uniformization of super-Riemann surfaces shall be used to study
the
superstring measure.  The super-Schottky group is generated by g
transformations
$T_n$, n=1,...,g, acting on the super-complex plane with coordinate
$Z=(z,\theta)$
$${{T_n(Z)-Z_{1n}}\over {T_n(Z)-Z_{2n}}}~=~K_n {{Z-Z_{1n}}\over {Z-Z_{2n}}}~~~~
\vert K_n\vert~<~1
\eqno(3)$$
where $Z_{1n}=(\xi_{1n}, \theta_{1n})$ and $Z_{2n}=(\xi_{2n}, \theta_{2n})$
are attractive and repulsive super-fixed points respectively.

It has been found that the measure on supermoduli space for the Neveu-Schwarz
sector of superstring theory, corresponding to the propagation of bosonic
states
in the loops, is simpler than that of the Ramond sector [7][8][9].  The
holomorphic part [10][11][12]
multiplied by the period-matrix factor is
$$\eqalign{{1\over {d {\cal V}_{ABC}}}~\prod_{n=1}^g~&{{dK_n}\over {K_n^{3\over
2}}}
{}~{{dZ_{1n} dZ_{2n}}\over {Z_{1n}-Z_{2n}}} \left({{1-K_n}\over 
{1-(-1)^{B_n} K_n^{1\over 2}}}\right)^2~[det(Im~{\cal T})]^{-5}
\cr
\cdot\prod_\alpha~^\prime \prod_{p=1}^\infty & 
\left({{1-(-1)^{N_\alpha^B} K_\alpha^{p-{1\over
2}}}
\over {1-K_\alpha^p}} \right)^{10}
~\prod_\alpha~^\prime \prod_{p=2}^\infty \left({{1-K_\alpha^p}
\over {1-(-1)^{N_\alpha^B} K_\alpha^{p-{1\over 2}}}}\right)^2
\cr}
\eqno(4)$$
where the infinitesimal super-projective invariant volume element is
$$\eqalign{d {\cal V}_{ABC}~&=~{{dZ_A dZ_B dZ_C}\over
{[(Z_A-Z_B)(Z_C-Z_A)(Z_B-Z_C)]^{1\over 2}}}\cdot {1\over {d\Theta_{ABC}}}
\cr
\Theta_{ABC}~&=~{{\theta_A(Z_B-Z_C)+
\theta_B(Z_C-Z_A)+\theta_C(Z_A-Z_B)+\theta_A
\theta_B\theta_C}\over {[(Z_A-Z_B)(Z_C-Z_A)(Z_B-Z_C)]^{1\over 2}}}
\cr}
\eqno(5)$$
and the super-period matrix is
$${\cal T}_{mn}~=~{1\over {2 \pi i}} \left[~ln~K_n~\delta_{mn}~+~
\sum_\alpha~^{(m,n)}~ln \left[{{Z_{1m}-V_\alpha Z_{1n}}\over {Z_{1m}-V_\alpha
Z_{2n}}} {{Z_{2m}- V_\alpha Z_{2n}}\over {Z_{2m}-V_\alpha Z_{1n}}} \right]
\right]
\eqno(6)$$
Selecting $B_n$ to be 0 or 1, depending on the boundary conditions around the
g B-cycles, produces $2^g$ spin structures associated with the exchange of 
bosonic states in the g loops.  The number $N_\alpha^B$ equals
$\sum_{n=1}^g~B_n N_\alpha^n$, where $N_\alpha^n$ is the number of times that
the generator $T_n$ or its inverse in the product $V_\alpha$.

Because the measure (4) does not contain the contribution of the Ramond sector,
it can only be used to obtain a partial estimate of the amplitude at large
genus.  However, the restriction to the Neveu-Schwarz sector just necessitates
a larger integration region corresponding to the different spin structures, but
this would only modify the bound by an exponential function of the genus,
as the number of spin structures increases as $2^{2g}$.   The choice of
spin structure is defined by the sign of the square root $K_n^{1\over 2}$
and $K_\alpha^{1\over 2}$ in
equation (4).
While a modular transformation may map the expression (4), with a given choice
of signs for the square roots $K_n^{1\over 2}$ for all n,
into a measure with different signs for the square roots, corresponding to
other spin structures, it will be demonstrated later that the integral would be altered only by an exponential
factor, from an analysis of sums of $\vert K_\alpha \vert$ over elements of
the Schottky group [4] and integration regions.

Certain advances have been made in obtaining the Ramond  contribution to the
full superstring measure.  The non-zero mode part of the most general amplitude
with an arbitrary number of Ramond loops $g_R$ and $g_{NS}$ Neveu-Schwarz loops
[13] is
$$\eqalign{\prod_{i=1}^{g_R} \prod_{n=2}^\infty &\left[{{1 \pm (K_i^R)^n}\over
{1- (K_i^R)^n}} \right]^8 \cdot \left[{{1 \pm K_i^R}\over {K_i^R}} \right]^{10}
\cr
&\cdot \prod_{\alpha, NS}~^\prime \prod_{n=2} \left[{{1-K_\alpha^{n-{1\over
2}}}
\over {1-K_\alpha^n}}\right]^8 \cdot \left[{{1-K_\alpha^{1\over 2}}\over
{1-K_\alpha}}\right]^{10}
\cr}
\eqno(7)$$
and the ghost zero-mode contribution from one Ramond loop is simply
$$4{{dK_R}\over {K_R}} {{dB_g}\over {B_g}} d \kappa_R {1\over {(1-\xi_{2g})}}
\eqno(8)$$
where $B_g$ is a multiplier variable defined in terms of the fixed points in a
subsequent formula and $\kappa_R$ is the fermionic modulus [13].

The overall $OSp(2 \vert 1)$ invariance can be used to fix two of the
super-fixed points
and the even coordinate of a third superfixed point, leaving 3g-3 even
moduli and 2g-2 odd moduli amongst the super-Schottky group parameters [7].
The integration region is defined to be the fundamental domain of the
super-mapping class group in super-Teichmuller space.  Equivalently, one may
use the intersection of the fundamental region of the super-modular group in
the space of positive-definite, symmetric  super-period matrices with the set
of ${\cal T}_{mn}$ associated with a super-Riemann surface, which will be
contained in the set of ${\cal T}_{mn}$ such that the ordinary period matrix
$\tau_{mn}$, the complex number-valued part of ${\cal T}_{mn}$,  lies in a
fundamental domain of the modular
group and corresponds to a Riemann surface.  This leads to an infinite number
of conditions on the multipliers $K_n$ and fixed points $\xi_{1n},~\xi_{2n}$ [4]
[14].

It has been shown in the analysis of the closed bosonic string that these
inequalities may be satisfied by certain categories of isometric circles,
$I_{T_n}=\{z\in {\hat {\Bbb C}}\vert \vert \gamma_n z+\delta_n\vert = 1\}$,
\hfil\break
$T_n z= {{\alpha_n z+\beta_n}\over {\gamma_n z +\delta_n}}$ which
would then represent a subset of moduli space.  In particular, the following
configurations of isometric circles
$$\eqalign{ (i) &~~{{\epsilon_0}\over {g^{1-2q^\prime}}}~\le~\vert K_n\vert~\le
{{\epsilon_0^\prime}\over {g^{1-2q^\prime}}}
{}~~~~~~\delta_0 \le \vert\xi_{1n}-\xi_{2n}\vert \le \delta_0^\prime~~~0\le
q^\prime \le {1\over 2}
\cr
(ii) &~~{{\epsilon_0}\over {g^{1-2q^\prime}}} \le \vert K_n\vert \le
{{\epsilon_0^\prime}\over {g^{1-2q^\prime}}}
{}~~~~~~~~~ {{\delta_0}\over {g^q}}\le \vert \xi_{1n}-\xi_{2n}\vert \le
{{\delta_0^\prime}\over {g^q}}~~~~~~0 < q \le q^\prime < {1\over 2}
\cr
(iii) &~~ \epsilon_0 \le \vert K_n\vert \le \epsilon_0^\prime
{}~~~~~~~~~~~~~~~~~~~~~ {{\delta_0}\over {\sqrt g}} \le \vert \xi_{1n}-\xi_{2n}
\vert \le {{\delta_0^\prime}\over {\sqrt g}}
\cr}
$$
describe a subset of Teichmuller space consistent with a cut-off on the radii
of the
isometric circles, or equivalently the size of the handles in the intrinsic
metric on the surface, $r_{I_{T_n}}^2=\vert \gamma_n\vert^{-2} \gtrsim {1\over
g}$ [15]. 

The absence of divergences in N-point superstring
amplitudes {\it at any given finite order} in perturbation theory implies that
the restriction on the minimum length of closed geodesics should be removed.
Effectively closed infinite-genus surfaces, besides completing the domain of
string perturbation theory, may have an essential role in the superstring
path integral.  Configurations of isometric circles arising in the super-Schottky uniformization of these surfaces shall then be included in the 
supermoduli space integral.
The ranges ${{\epsilon_0}\over {n^{q^{\prime\prime}}}} \le \vert K_n\vert
\le {{\epsilon_0^\prime}\over {n^{q^{\prime\prime}}}},~n=1,...,g,~~
 q^{\prime\prime}>1$, and
${{\delta_0}\over {n^{q^{\prime\prime\prime}}}} \le \vert \xi_{1n}-\xi_{2n}
\vert \le {{\delta_0^\prime}\over
{n^{q^{\prime\prime\prime}}}},~n=1,...,g,~~q^{\prime\prime\prime}>{1\over 2}$,
therefore will be considered also.
While the parameters $q,~q^\prime,~q^{\prime\prime}$ and
$q^{\prime\prime\prime}$ are initially chosen to be continuous parameters,
to avoid overcounting in the path integral, it is necessary to select a
discrete
set of values. Specifically, it is sufficient to sum over the values
$$\eqalign{&q ~\le~ q^\prime_N ~=~N {{ln~ \left({{\epsilon_0^\prime}\over
{\epsilon_0}}\right)}\over {2~ln~g}}~~~~~~~N~=~0,~1,...,\left[ {{ln~g}\over
{ln~\left({{\epsilon_0^\prime}\over {\epsilon_0}}\right)}}\right]
\cr
&q^{\prime\prime}_{\tilde N}~=~1~+~{\tilde
N}{{ln~\left({{\epsilon_0^\prime}\over {\epsilon_0}}
\right)}\over {ln~n}}~~~~~~~{\tilde N}~=~1,~2, ...
\cr}
\eqno(9)$$
with the values of $q^{\prime\prime\prime}$ specified later.  The values of N
follow from the non-overlapping of the ranges  ${{\epsilon_0}\over
{g^{1-2q^\prime}}}\le \vert K_n\vert
\le {{\epsilon_0^\prime}\over {g^{1-2q^\prime}}}$ for different choices of
$q_N^\prime$ and the covering of the interval $0 \le q \le q^\prime_N < {1\over
2}$
[4].  Similarly, non-overlapping of the ranges ${{\epsilon_0}\over
{n^{q^{\prime
\prime}}}} \le \vert K_n \vert \le {{\epsilon_0^\prime}\over
{n^{q^{\prime\prime}}}}$ requires that the set $\{q^{\prime\prime}_{\tilde
N}\}$ be
selected so that
$$\eqalign{ {{\epsilon_0^\prime}\over {n^{q^{\prime\prime}_{{\tilde N} +1}}}}
            ~&=~{{\epsilon_0}\over {n^{q^{\prime\prime}_{\tilde N}}}}
\cr
q^{\prime\prime}_{{\tilde N}+1}~-~q^{\prime\prime}_{\tilde N}~&=~{{ln \left(
{{\epsilon_0^\prime}\over {\epsilon_0}}\right)}\over {ln~n}}
\cr}
\eqno(10)$$
and the last equality is satisfied by the sequence
$\{q^{\prime\prime}_{\tilde N}\}$ in equation (9).

An estimate for the superstring path integral with measure (4) will now be
given based on surfaces near the degeneration locus.
The use of this measure initially reduces the divergence from
$\int{{d \vert K\vert}\over {\vert K\vert^3 (log \vert K\vert)^{13}}}$ for
bosonic strings to $\int {{d \vert K\vert}\over {\vert K\vert^2 (log \vert
K\vert)^5}}$ for the Neveu-Schwarz string as $\vert K\vert \to 0$, reflecting
the existence of a tachyon in both cases and a shift in the value of the
square of its momentum [16].  It has been noted, however, that an extra factor
of $K^{1\over 2}$ in the holomorphic part of the measure arises for Ramond
fermions circulating in the loops and also when the GSO projection is applied
to the Neveu-Schwarz sector.  A sum over spin structures, weighted with a phase
factor $(-1)^{B_n}$, introduces the factors $K_n^{1\over 2},~n~=~1,...,~g$,
since
$${1\over {[1-K_n^{1\over 2}]^2}}~-~{1\over {[1+K_n^{1\over 2}]^2}}
{}~=~{{4K_n^{1\over 2}}\over {[1-K_n]^2}}
\eqno(11)$$
signalling the absence of a tachyon singularity.

While restriction to the Neveu-Schwarz sector leads to a partial estimate of
the amplitude, an estimate of the full amplitude, including a sum over all $2^{2g}$
spin structures, can be obtained by considering an integration region
consisting
of more than one fundamental domain of the modular group.

This may be seen more concretely at one loop.  There the fundamental region of
the modular group $SL(2;{\Bbb Z})$ is mapped onto the adjoining fundamental
region by the shift $\tau \to \tau + 1$.  Labelling the spin structures by the
sign of the world-sheet fermions upon one traversal of the A- and B-cycles,
there is one odd spin structure (++) and three even spin structures $(+-),
(-+)$ and $(--)$ at genus 1.  Under the map $\tau \to \tau +1$, $(++) \to (++)$
while $(+-) \to (+-)$, $(-+) \to (--)$ and $(--) \to (-+)$.  Similarly, the
modular transformation $\tau \to -{1\over \tau}$, which maps the original
fundamental region to another copy in the upper half plane, leaves (++)
invariant, whereas $(+-) \to (-+)$, $(-+) \to (+-)$ and $(--) \to (--)$.
Given one odd spin structure and one even spin structure, all four spin
structures may be obtained by using the two transformations $\tau \to \tau +1$,
$\tau \to -{1\over \tau}$ and their product.  It may be recalled that the
sum over all four spin structures is required for the vanishing of the
one-loop partition function
$$Z_1(\tau)~=~{1\over 2} {1\over {\eta(\tau)^4}} [ \eta_{(++)}
\theta_1^4 (0 \vert \tau)~+~\eta_{(+-)} \theta_2^4(0 \vert \tau)~+~
\eta_{(-+)} \theta_4^4(0\vert \tau)~+~\eta_{(--)} \theta_3^4 (0 \vert \tau)]
\eqno(12)$$
with the choice of phases  $\eta_{(++)}~=~\pm 1,~\eta_{(+-)}~=~-1,
\eta_{(-+)}~=~-1$ and $\eta_{(--)}~=~1$
\hfil\break[17].

At genus g, there are $2^{g-1}(2^g-1)$ odd and $2^{g-1}(2^g+1)$ even spin
structures.  Thus, the problem of estimating the full superstring amplitude
becomes one of summing over all $2^{2g}$ spin structures, beginning with the
set of $2^g$ spin structures associated with the Neveu-Schwarz sector of the
string.  Since the number of Dirac zero modes modulo 2 on a Riemann surface is
a modular invariant, the sets of odd spin structures and even spin structures
each
form a representation of the modular group and all $2^{2g}$ spin structures
arise from modular transformations of one odd and one even spin structure.
Thus, if the set of $2^g$ spin structures did contain at least one odd and
one even spin structure, summation over all $2^{2g}$ spin structures can be
achieved by enlarging the integration region to contain those regions that
are images of the original fundamental region under the corresponding
modular transformations.

The choice of signs for spin structures at higher genus can be deduced by
noting that the number of Dirac zero modes modulo 2 is additive when two
Riemann surfaces are glued [18].  Thus an odd spin structure at genus 2 is
composed of an odd spin structure at genus 1 and an even spin
structure at genus 1, while an even spin structure at genus 2 consists of a
sum of two even spin structures at genus 1 or two odd spin structures at genus
1.  The odd spin structures at genus 2 may be listed as $(+++-), (++-+),
(++--), (+-++), (-+++)$ and $(--++)$ while the even spin structures are
$(++++), (+-+-), (+--+), (+---), (-++-), (-+-+), (-+--), (--+-), (---+)$ and
$(----)$.  The Neveu-Schwarz sector, defined by anti-periodic boundary
conditions
around the A-cycles for the world-sheet fermions, corresponds to the spin
structures $(-+-+), (-+--), (---+)$, and $(----)$, which are all even.  Since
the
addition of even spin structures is again even, the Neveu-Schwarz sector should
always consist of even spin structures.  This implies that modular
transformations of the Neveu-Schwarz sector initially can only span the set of
even spin structures.

This difficulty of summing over the odd spin structures
can be resolved by enlarging the set of $2^g$ spin structures to include
an odd spin structure.  This can be done as follows.
Considering the map from the parallelogram to the annulus, the target space
coordinate is $X~=~e^{2 \pi i z}~=~x~+~ \theta \psi$, under the transformation
$z \to z+1$,  $X \to X$, but the fermionic coordinates change sign
$\theta \to e^{\pi i} \theta$ and $\psi \to e^{\pi i} \psi$.
Since the annulus can then be mapped to the Schottky plane with two disks
removed without changing the sign of the fermions, a change in the
transformation of the Grassmann coordinate $\theta$ will induce a change in the
sign of the world-sheet fermion leading to an odd spin structure.

While the choice of phase is required for the transformation of amplitudes
corresponding to distinct spin structures, the absolute values of the
amplitudes are also mapped into each other.  Consequently, modular invariance
is not only a property of the standard sum over spin structures
with the correct phases in the superstring amplitude, but it is also valid for
a sum of absolute values of amplitudes associated with different sets of
spin structures.  Specifically, if $\eta_{(i)}$ denotes the phase associated
with the spin structure (i) and $\sigma_r$ is the modular transformation,
then the superstring amplitude
$A_{N,g}~=~\int_{F_g}~ I_{N,g}$, where $F_g$ is the fundamental region of the
modular group, is invariant so that
$$\int_{F_g} \sum_{(i)=1}^{2^{2g}}~\eta_{(i)}
I_{N,g}^{(i)}~=~\int_{\sigma_r(F_g)}\sum_{(i)=1}^{2^{2g}}~
\eta_{\sigma_r(i)}~I_{N,g}^{\sigma_r(i)}
\eqno(13)$$

This sum can be arranged into sets, each consisting of $2^g$ spin structures,
so that the sum over spin structures within any individual set introduces a
factor of $K_n^{1\over 2}$, in the same manner as the sum over spin structures
(11) in the Neveu-Schwarz sector, implying removal of the tachyon singularity.
Denoting the sets as $S_r,~r=1,2,3,...$, where $S_1$ represents the
Neveu-Schwarz sector, it can be shown that there are $2^{g-1}$ odd spin
structures and $2^{g-1}$ even spin structures in each set $S_r,~r=2,3,...$.
Recalling that modular transformations acting on the Neveu-Schwarz sector
will generate the other even spin structures, but no odd spin structures,
it can be seen that $2^{g-1} \cdot 2^g$ even spin structures naturally arise in
this way, leaving $2^{g-1}$ spin structures.  To constitute a set of $2^g$
spin structures, $2^{g-1}$ odd spin structures are needed.  An examination of
the signs of the associated with the B-cycles reveals that it is this set of
$2^{g-1}$ odd and $2^{g-1}$ even spin structures which allows for cancellation
of the tachyon divergence, rather than the sets of $2^g$ even spin structures,
other than the Neveu-Schwarz sector, that belong to the group $\{S_r\}$.
Beginning with this set, which may be labelled as $S_2$, modular
transformations might then be used to generate the remaining $2^{g-1}(2^g - 2)$ odd spin structures and simultaneously $2^{g-1}(2^g-2)$ even spin structures, so that the sets $S_r,~r=3,4,...$ contain an equal number of odd and even spin
structures. This description of the sets $\{S_r\}$ can be verified at arbitrary genus. This labelling can be done for genus 1, 2 and 3 in particular.

The existence of a modular transformation mapping the set $S_2$ into $S_r,~
r=3,4,...$ should follow from the absence of the tachyon singularity for
each of the sets.  The tachyon divergences in the limit
$\vert K_n\vert \to 0$ cancel only when the correct choice of spin structures,
consistent with modular transformations of the integrand corresponding
to $S_2$, is used.  One can consider the following sum over spin structures
$$\eqalign{&\int_{F_g}~\sum_{(a_{NS})=1}^{2^g}~\eta_{(a_{NS})}
I_{N,g}^{(a_{NS})}~+~\int_{F_g}~\sum_{(a_e)=1}^{2^{g-1}}~\eta_{(a_e)}
I_{N,g}^{(a_e)}
\cr
&~~~+~\int_{F_g}~\sum_{(a_o)=1}^{2^{g-1}}~\eta_{(a_o)} I_{N,g}^{(a_0)}
{}~+~\sum_{r=3}^{2^g}~\int_{F_g}~\sum_{(a_e)=1}^{2^{g-1}}~
\eta_{\sigma_r(a_e)}~I_{N,g}^{\sigma_r(a_e)}
\cr
&~~~+~
\sum_{r=3}~\int_{F_g}~\sum_{(a_o)=1}^{2^{g-1}}~
\eta_{\sigma_r(a_o)}~I_{N,g}^{\sigma_r(a_o)}
\cr}
\eqno(14)$$
where $S_2~=~\{(a_e)\} \cup \{(a_o)\}$, with $\{(a_e)\}$ representing even
spin structures and $\{(a_o)\}$ representing odd spin structures, and
$S_r~=~\{\sigma_r(a_e)\}
\cup \{\sigma_r(a_o)\}$.   The magnitude of this sum can then be
bounded since
$$\left \vert \int_{F_g}~\sum_{(i)=1}^{2^{2g}}~\eta_{(i)} I_{N,g}^{(i)}
\right\vert~\le~\sum_{r=1}~\left\vert~\int_{F_g}~\sum_{(a) \in S_r}
\eta_{(a)} I_{N,g}^{(a)} \right\vert
\eqno(15)$$

Each of the absolute values of the integrals, labelled by r, is free of the
infrared divergence in the $\vert K_n\vert \to 0$ limit, since the sum over
the $2^g$ spin structures in $S_r$ removes the tachyon singularity.
This can be verified at genus 1, for example, as the infinities associated
with the Neveu-Schwarz spin structures $(-+)$ and $(--)$ in the limit $\tau \to
i \infty$ cancel
$$lim_{\tau \to i \infty}~ \left[{{\theta_3^4(0 \vert \tau)}
\over {\eta(\tau)^4}}~-~{{\theta_4^4(0\vert \tau)}\over {\eta(\tau)^4}}
\right]
{}~=~0
\eqno(16)$$
It will be shown in the following analysis that the absence of divergences
in the limit $\vert K_n\vert \to 0$ is used in the proof of finiteness of the
amplitudes in the other degeneration limits.  This implies that when the
divergences in these limits are eliminated, bounds on the magnitudes of the
integrals $\int_{F_g}~\sum_{a \in S_r}~\eta_{(a)} I_{N,g}^{(a)}$ will suffice
to provide an upper bound on the total superstring amplitude
$\int_{F_g}~\sum_{(i)=1}^{2^{2g}}~\eta_{(i)} I_{N,g}^{(i)}$.

Another divergence associated with the coincidence of fixed points $\xi_{1n}$
and $\xi_{2n}$ arises in the Neveu-Schwarz sector, even after summing over
spin structures.  One might therefore wish to use the combination
of the Neveu-Schwarz measure with the part of the Ramond contribution in
equations (7) and (8) to estimate the genus-dependence of the amplitudes.
For example, the ghost zero-mode contribution at one Ramond loop contains
a dependence on the multiplier variable $B_g$ which represents a softening
of the divergences in the $\vert B_g\vert \to 0$ limit.

The actual finiteness of the superstring amplitudes demonstrates that use of
the entire superstring measure would remove this divergence.  The analysis of
[6]
shows that this is achieved by transferring the fixed-point divergence to
an integral  associated with the multipliers $K_n$, which has been shown
to be finite in the limit $\vert K_n\vert$ by the above argument.
Specifically,
the fixed points can be expressed in terms of alternative variables $B_m$
and $H_m$ [11] which may be defined in terms of different radii in the
representation of the N-punctured genus-g surface as the sewing together
of 2g-2+N three-punctured spheres (Fig. 1)

\input epsf.tex

\vbox{
\epsfysize=2.4in
\centerline{\epsfbox{puncture.eps}}
\vskip 0.2in
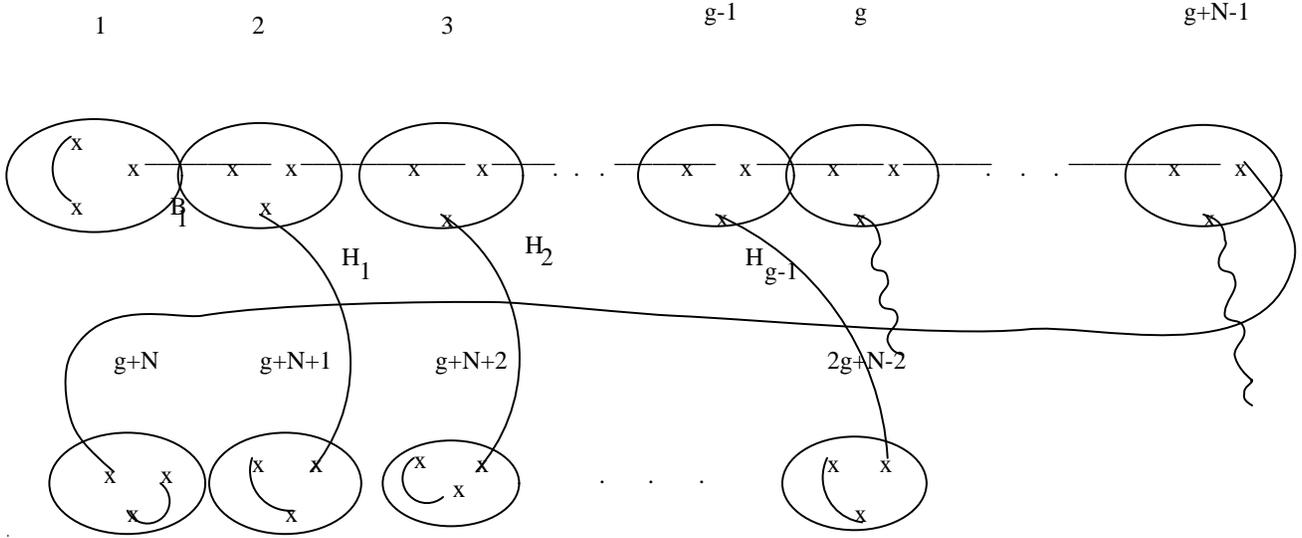
\noindent{\bf Fig. 1. An N-punctured genus-g surface can be constructed using
2g-2+N three-punctured spheres.}}

In the twistor-string formalism, the amplitude contains the picture-changing
operators $F^{\pm}~=~[Q, \xi^{\pm}]$, where Q is a BRST operator derived from
the energy-momentum tensor and ghosts and $\xi^{\pm}$ are spin-0 fermions
required in the expression for the bosonized ghosts.  The locations of the
picture-changing operators are chosen to be arbitrary points on the 2g-2+N
spheres.  Since a change in the location of $F^{\pm}$ implies that
$[Q, \xi]~\to~[Q, \xi]~+~ \int dz [Q, \partial_z \xi]$, the contour of Q
initially surrounds the curve joining the initial and final positions of the
picture-changing operators [19].  In the degeneration limit $H_m \to 0$, where
the Riemann surface splits into two components of genus g-1 and 1, the contour
of Q can be pulled off the picture-changing operator to surround the three
punctures of $S_{i+g+N}$ (Fig. 2).

\input epsf.tex

\vbox{
\epsfysize=2.4in
\centerline{\epsfbox{contour.eps}}
\vskip 0.2in
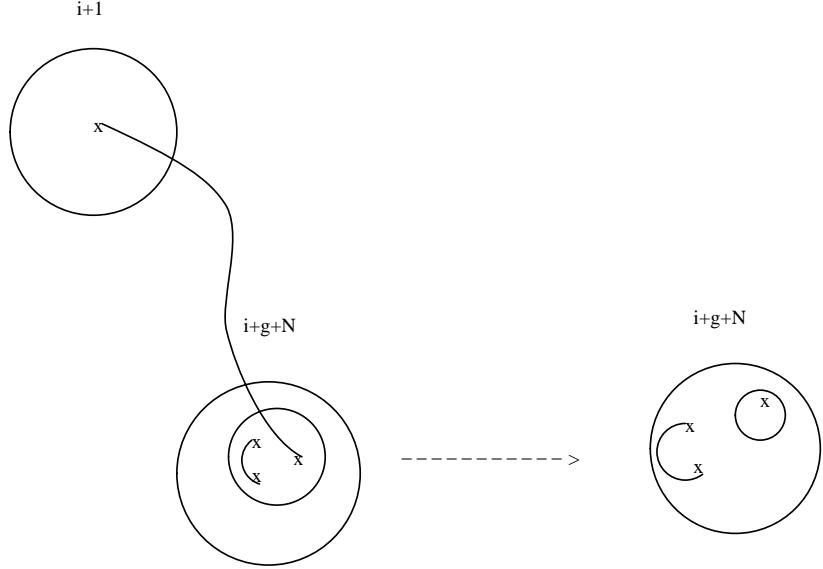
\noindent{\bf Fig. 2. The contour of Q is pulled past two of the punctures
sewed together to form a handle on $S_{i+g+N}$ to surround the third
puncture.}}

Around each of the sewed punctures is a closed loop $C_i$ or radius $R_i$
and the contribution from each of the Beltrami differentials is
$$\left\vert \prod_{i=1}^{3g-3+N}~\int_{C_i} {{y_i b(y_i)}\over {R_i}} dy_i
\right\vert^2
\eqno(17)$$
Around the first puncture is the contour integral involving the Beltrami
differential for $K_n$, while at the third puncture is the operator
$$\vert c~exp(-\phi^+ - \phi^-) exp(h^+ +h^-) \psi^+ \psi^- \vert^2
\eqno(18)$$
where c is a right-moving boson of spin-${1\over 2}$, $\phi^{\pm}$ are two
scalar bosons of screening charge 2 defining the bosonized ghosts, $h^{\pm}$
are a pair of right-moving spin-0 fermions [6].  Anticommuting Q with the
Beltrami differential for $K_n$ produces a derivative with respect to $K_n$.
This total derivative does not represent a divergence because the amplitude
already has been shown to be finite at the boundary $K_n~=~0$.  Anticommutation
of Q with the operator at the third puncture produces no terms with zero-modes
of the spin-0 fermions $\psi^{\pm}$, and therefore the amplitude is independent
of the locations of the picture-changing operators.  Since all of the picture-
changing operators may be moved to the third puncture, the zero-modes of
$\psi^+$ and $\psi^-$ cancel, implying that the amplitude associated with the
genus 1 component of the Riemann surface vanishes [6].  This leaves the
contribution from the boundary of the $K_i$ region and, in this sense, it represents a transfer from an integral involving $H_m$ to a finite integral
involving $K_n$.  Consequently, divergences are absent in the (Type IIB
Green-Schwarz) superstring amplitudes in the degeneration limit $H_m\to 0$.
A similar argument may be used to demonstrate that the amplitudes are also 
finite in the $B_m$ limit [6].

This finiteness result, valid at any given order, suggests a genus-dependence
for the superstring amplitudes mentioned in the introduction.  As the
amplitude is finite in each of the degeneration limits $K_n,~H_m,~B_m \to 0$,
and the leading-order behaviour of large-order amplitudes is essentially
determined by taking these limits simultaneously, it follows that the products
of the integrals over these multipliers should be bounded by an exponential
function of the genus, $B_K^g \cdot B_H^{g-2} \cdot B_B^{g-1}$, where
$B_K,~B_H$ and $B_B$ are upper bounds for each of the $K_n, H_m$ and $B_m$
integrals respectively.

However, any conclusion derived from the boundedness of the amplitudes in
each of the degeneration limits can be improved by a more precise estimate
of the genus-dependence at large orders.  Ideally, therefore, one would like
to begin the present calculation of the amplitudes with the entire superstring
measure. The argument of [6],
transferring the problem of the fixed-point divergence to the multiplier
integral, implies that one can proceed with the estimate of the amplitudes
using the measure for the Neveu-Schwarz sector and the measure for the Ramond
sector, with the appropriate sum over spin structures, and apply modular
transformations to obtain bounds for the sum over all $2^{2g}$ spin structures. 
This procedure would be essentially
equivalent to setting an upper bound for an amplitude using the hypothesized divergence-free superstring
measure in the super-Schottky paramatrization.
Moreover, in terms of the sums over spin structures,
$$\eqalign{\left\vert\int_{F_g}~\sum_{(a) \in
S_r}~\eta_{(a)}~I_{N,g}^{(a)}\right\vert
{}~&=~ \left\vert~\int_{F_g}~\sum_{(a_e)=1}^{2^{g-1}}~\eta_{\sigma_r(a_e)}
{}~I_{N,g}^{\sigma_r(a_e)}~+~\int_{F_g}~\sum_{(a_o)=1}^{2^{g-1}}~
\eta_{\sigma_r(a_0)}~I_{N,g}^{\sigma_r(a_o)}\right\vert
\cr
{}~&=~\left\vert~\int_{\sigma_r^{-1}(F_g)}~\sum_{(a_e)=1}^{2^{g-1}}~
\eta_{(a_e)}~I_{N,g}^{(a_e)}~+~\int_{\sigma_r^{-1}(F_g)}~
\sum_{(a_0)=1}^{2^{g-1}}~\eta_{(a_0)}~I_{N,g}^{(a_0)}\right\vert
\cr}
\eqno(19)$$
Since the integral involving the spin structures $\{\sigma_r(a_e)\}\cup
\{\sigma_r(a_o)\}$ has been converted to an integral involving the spin
structures
$\{(a_e)\}\cup \{(a_o)\}~=~S_2$, with a different integration region, it
is sufficient to consider just the spin structures in $S_1$ and $S_2$ and
to determine the effect of integrating over a different region in Schottky
group parameter space.  Upon evaluation of the integrals over the ranges for
the parameters $K_n, \xi_{1n}, \xi_{2n}$ specified earlier, it follows that the
shift of the integration region from $F_g$ to $\sigma_r^{-1}(F_g)$ will
change the bound on the absolute values of the integrals by an
exponential factor $\sigma(r)^g$.  Let $\sigma_{max}~=~max_r~\sigma(r)$.
Then
$$\left\vert
\int_{F_g}~\sum_{(i)=1}^{2^{2g}}~\eta_{(i)}~I_{N,g}^{(i)}\right\vert~\le~(k~
\sigma_{max})^g~\sum_{r=1,2}~\left\vert~\int_{F_g}~\sum_{(a)\in S_r}~
\eta_{(a)}~I_{N,g}^{(a)}\right\vert
\eqno(20)$$
for some positive integer k.
As mentioned earlier, the spin structures in $S_2$ can be obtained by changing
the boundary conditions for world-sheet fermions at one of the A-cycles.
Since this seems to be equivalent to changing the boundary condition for a
single
Grassmann coordinate, the bounds on the integral corresponding to the
set $S_2$ should differ from those for $S_1$ only by an exponential
factor of the genus.
It therefore follows from equation (20) that the
estimates of the amplitudes given below should have the same genus-dependence
as those for the full superstring measure.

Combining the holomorphic part of the measure with its complex conjugate gives
a multiplier integral with a dominant part
$\int{{d\vert K_n\vert}\over {\vert K_n\vert \left(ln \left({1\over {\vert
K_n\vert}}\right)\right)^5}}~=~ {1\over 4} \left[ln \left({1\over {\vert
K_n\vert}}\right)\right]^{-4}$
in the limit $\vert K_n\vert \to 0$.  With the limits in categories (i) and
(ii),
the integral gives a factor $((1-2q^\prime) ln~g)^{-5} ln
\left({{\epsilon_0^\prime}\over {\epsilon_0}}\right)+O(((1-2q^\prime)
ln~g)^{-6})$.  For the third category of isometric circles, a factor of
${1\over 4}\left[\left(ln ~{1\over {\epsilon_0^\prime}}\right)^{-4}
-\left(ln~{1\over {\epsilon_0}}\right)^{-4}\right]$ is obtained.
The contribution of a configuration of isometric circles with $N_0$ circles
belonging to either of the first two categories and $g-N_0$ circles in the
third category to the multiplier integral is
$$\eqalign{(1-2 {\bar q}^\prime)^{N_0}& ln \left({{\epsilon_0^\prime}\over
{\epsilon_0}}\right)^{N_0} (ln~g)^{-5N_0}
\cr
\cdot\biggl[&{1\over 4}ln \left({{\epsilon_0^\prime}\over {\epsilon_0}}\right)
\left(ln~{1\over {\epsilon_0}}\right)^{-4} \left(ln~{1\over
{\epsilon_0^\prime}}\right)^{-4}
\cr
 &~\left(ln~{1\over {\epsilon_0}}+ln~{1\over {\epsilon_0^\prime}}\right)
\left(\left(ln~{1\over {\epsilon_0}}\right)^2 +\left(ln~
{1\over {\epsilon_0^\prime}}\right)^2\right) \biggr]^{g-N_0}
\cr
(1-2 {\bar q}^\prime)^{N_0}~&=~ \prod_{m=1}^{N_0}~(1-2q_m^\prime)
\cr}
\eqno(21)$$

In this formula, ${\bar q}^\prime$ represents a weighted average of the values
$q^\prime_m$ associated with the $N_0$ isometric circles in the first two
categories.
There are
$${{\left(N_0 + {{(ln~g)^2}\over {2 \left(ln \left({{\epsilon_0^\prime}\over
{\epsilon_0}}\right)\right)^2}} -1\right)!}
\over { N_0! ~\left({{(ln~g)^2}\over {2 \left(ln \left({{\epsilon_0^\prime}
\over {\epsilon_0}}\right)\right)^2}} - 1\right)!}}
\eqno(22)$$
different partitions of the $N_0$ circles into the ${{(ln~g)^2}
\over {2 \left(ln\left({{\epsilon_0^\prime}\over
{\epsilon_0}}\right)\right)^2}}$
subcategories labelled by the indices $q$ and $q^\prime$, and each partition
$\{n_i\}$ is weighted by a factor ${1\over {n_1!...n_r!}}$, with
$n_1+...+n_r=N_0$, since one set of inequalities defining the fundamental
domain of the modular group, $(Im~\tau)_{ss} \ge (Im~\tau)_{rr},~s \ge r$, must
be
satisfied for each subcategory, leading to restrictions on the ranges of the
multipliers [4].

Instead of integrating over the super-fixed points, it is useful to use another
set of coordinates on supermoduli space $\{K_n, B_m, H_m, {\cal \theta}_{1i},
{\cal \theta}_{2i}\}$ for which the holomorphic part of the integral over the
super-fixed points [11] is replaced by
$$\prod_{m=2}^g~{{dB_m}\over {B_m^{3\over 2}}}~\prod_{m=2}^{g-1} {{dH_m}\over
{H_m^{3\over 2}}}~\prod_{i=2}^{g-1}~
d{\vartheta}_{1i}~\prod_{i=1}^g~d{\vartheta}_{2i}
\eqno(23)$$
where
$$\eqalign{\xi_{2n}~&=~\prod_{j=2}^n~B_j~~~{\theta}_{2n}={1\over {\sqrt
{\xi_{2n}}}} \vartheta_{2n}~~~ \theta_{2g}={1\over {\sqrt
{\xi_{2g}}}}{\vartheta_{2g}}
\cr
\xi_{1n}~&=~{{\xi_{2n}}\over {1-H_n-{\sqrt {H_n}} {\vartheta}_{1n}
{\vartheta}_{2n}}}~~~\theta_{1n}={{{\sqrt
{H_n\xi_{2n}}}{\vartheta}_{1n}+\theta_{2n}}\over {1-H_n}}
\cr}
\eqno(24)$$
The relation involving $H_n$ can be inverted to give
$$\eqalign{H_n~&=~{{\xi_{1n}-\xi_{2n}}\over
{\xi_{1n}}}-{{({\xi_{1n}-\xi_{2n}})^{1\over 2}}\over {{\xi_{1n}}^{1\over 2}}}
\vartheta_{1n} \vartheta_{2n}
\cr
\vert H_n\vert^{-1}~&=~\left\vert {{\xi_{1n}}\over
{\xi_{1n}-\xi_{2n}}}\right\vert
\biggl[1+{1\over 2} {{\xi_{1n}^{1\over 2}}\over {(\xi_{1n}-\xi_{2n})^{1\over
2}}} \vartheta_{1n} \vartheta_{2n}
\cr
&~~~~~~~~~~~~~~~~~~~~~~~+{1\over 2} {{{\bar \xi}_{1n}^{1\over 2}}\over
{{\overline {(\xi_{1n}-\xi_{2n})}}^{1\over 2}}}  {\bar \vartheta}_{1n} {\bar
\vartheta}_{2n}
\cr
&~~~~~~~~~~~~~~~~~~~~~~~+{1\over 4} {{\vert \xi_{1n}\vert}\over {\vert
\xi_{1n}-\xi_{2n}\vert}}
\vartheta_{1n} \vartheta_{2n}
{\bar \vartheta}_{1n} {\bar \vartheta}_{2n}\biggr]
\cr}
\eqno(25)$$
The dominant contribution to the integral over $\vert H_n\vert$
in N-point superstring scattering amplitudes with $N~\ge~4$ produces a
dependence on the fixed point distance of $\xi_{1n}-\xi_{2n}\vert^{-2}$.  This may be verified using the super-fixed points
$Z_{1n}$ and $Z_{2n}$.

For the parameters associated with the $g-N_0$ remaining isometric circles
in the third category, the constraints $(Im~\tau)_{ss} \ge (Im~\tau)_{rr},~s
\ge r$, lead to a reduction of the integral by a factor of $(g-N_0)!$.  After
summing over the possible values of $N_0$, the combination of the integrals
over the multipliers $K_n$ and $H_n$ with the weighting factors is bounded
above by
$$\eqalign{\sum_{N_0=0}^g~\sum_{{\{n_i\}}\atop {\sum_i n_i=N_0}}& {1\over {n_1!
... n_r!}}
(1-2{\bar q}^\prime)^{N_0}\left[ln~{{\epsilon_0^\prime}\over
 {\epsilon_0}}\right]^{N_0} (ln~g)^{-5N_0} g^{2 \sum_i q_i n_i}
\cr
\cdot\biggl[{1\over 4} ln~{{\epsilon_0^\prime}\over {\epsilon_0}}&  \left(ln~
{1\over {\epsilon_0^\prime}}\right)^{-4}\left(ln~{1\over
{\epsilon_0}}\right)^{-4}
\left(ln~{1\over {\epsilon_0}}+ln~{1\over {\epsilon_0^\prime}}\right)
\left(\left(ln~{1\over {\epsilon_0}}\right)^2+\left(ln~
{1\over {\epsilon_0^\prime}}\right)^2 \right) \biggr]^{g-N_0}
\cr
&~{{g^{(g-N_0)}}\over {(g-N_0)!}}
\cr}
\eqno(26)$$

The combinatorial factor ${{{g^{2\sum_i q_i n_i}}g^{g-N_0}}\over {n_1!...n_r!
(g-N_0)!}}$ obtains a maximum value when r=1 and $N_0={g\over 2}$.  If all of
the $N_0$ isometric circles are in the subcategory defined by the value
\hfil\break
$q_i~=~{1\over 2}-{{ln~{{\epsilon_0^\prime}\over {\epsilon_0}}}\over {ln~g}}$,
then the factor is approximately equal to $\left({{2\cdot e\cdot
\epsilon_0}\over {\epsilon_0^\prime}}\right)^g$.

The bound in equation (26) includes a division by a factor of $(g-N_0)!$ as a
result of the constraints on the imaginary part of the period matrix. For
isometric circles in the third category, it had been shown in [4] that there is
a sequential ordering of the diagonal elements of $Im~\tau$ in a local
neighbourhood of J circles upon restriction of arguments of the fixed-point
intervals $\xi_{1n}-\xi_{2n}$ and multipliers $K_n$, leading to a reduction
of the integral by an exponential function of the genus, and then, that the
period matrix inequalities for representative circles from each of the
${{(g-N_0)}\over J}$ local neighbourhoods by a factor of ${{(g-N_0)}\over J}!$.
However, as the genus increases, the values of ${Im~\tau}_{nn}$ for the circles
in each of the local neighbourhoods overlap, and the inequalities obeyed by the
representative circles will no longer be sufficient to imply the inequalities
for the entire set of $(g-N_0)$ circles.  In the large-genus limit, conditions
on the absolute values of the multipliers will again have to be imposed,
leading
to a further reduction of the integral.  Alternatively, one may state that J
tends to 1 in the large-genus limit.

The division by $(g-N_0)!$ rather than
 ${{(g-N_0)}\over J}!$ leads to an upper bound depending
exponentially on the genus for all $N_0$, as the combinatorial factor
is less than
$$\eqalign{g^{N_0}&~~~~~when~~~~~N_0~\le {{(ln~g)^2}\over
{2\left(ln~{{\epsilon_0^\prime}\over {\epsilon_0}}\right)^2}}
\cr
e^{g\left(1-2{{ln\left({{(ln~g)}\over {ln~{{\epsilon_0^\prime}\over
{\epsilon_0}}}}\right)}
\over {ln~g}}\right)}&~~~~~when~~~~~{{(ln~g)^2}\over
{2\left(ln~\left({{\epsilon_0^\prime}\over {\epsilon_0}}\right)^2\right)}}~\le~
{g\over {ln~g}}
\cr
\left[\left(1+{{ln~\lambda}\over \lambda}+{1\over \lambda}\right)\cdot
e\right]^g \cdot \left({{(ln~g)^2}\over {2\left(ln~{{\epsilon_0^\prime}\over
{\epsilon_0}}\right)^2}}\right)!&
{}~~~~~when~~~~~{g\over {ln~g}}~\le~N_0~\le~{g\over \lambda},~\lambda \gg 1
\cr
\left({{2 \cdot e\cdot \epsilon_0}\over {\epsilon_0^\prime}}\right)^g&~~~~~when
{}~~~~~{g\over \lambda}~\le~N_0~ \le~ {g\over 2}
\cr}
\eqno(27)$$
Each of these upper bounds must be multiplied by the number of partitions (22)
for each value of $N_0$, and since this is less than $(ln~g)^{2N_0}$, the
sum (26) is bounded above by
$$\eqalign{&{1\over {4^g}}~\left[\left(ln~{1\over
 {\epsilon_0^\prime}}\right)^{-4}
-\left(ln~{1\over {\epsilon_0}}\right)^{-4}\right]^g
{}~\left[\left(1+{{ln~\lambda}\over \lambda}+{1\over \lambda}\right)
\cdot e\right]^g \cdot \left({{(ln~g)^2}\over {2\left(ln~{{\epsilon_0^\prime}
\over {\epsilon_0}}\right)^2}}\right)!
\cr
&\biggl[1-4(1-2 {\bar q^\prime}) \left(ln~{1\over {\epsilon_0}}\right)^4
\left(ln~{1\over {\epsilon_0^\prime}}\right)^4
\cr
&~~~~~~~~~~~~~~~~~~~~~~~~~\cdot\left(ln~{1\over {\epsilon_0}}
+ln~{1\over {\epsilon_0^\prime}}\right)^{-1} \left(\left(ln~{1\over
{\epsilon_0}}
\right)^2+\left(ln~{1\over
{\epsilon_0^\prime}}\right)^2\right)^{-1}~(ln~g)^{-3}
\biggr]^{-1}
\cr
&~+~{1\over {4^g}}~\left[\left(ln~{1\over {\epsilon_0^\prime}}\right)^{-4}
-\left(ln~{1\over {\epsilon_0}}\right)^{-4}\right]^g
{}~\left({{2 \cdot e \cdot \epsilon_0}\over {\epsilon_0^\prime}}\right)^g
{}~g(1-{1\over \lambda})
\cr
&\biggl[4(1-2{\bar q^\prime}) \left(ln~{1\over {\epsilon_0}}\right)^4
\left(ln~{1\over {\epsilon_0^\prime}}\right)^4~\left(ln~{1\over {\epsilon_0}}
+ln~{1\over {\epsilon_0^\prime}}\right)^{-1}~\left(\left(ln~{1\over
{\epsilon_0}}\right)^2 +\left(ln~{1\over
{\epsilon_0^\prime}}\right)^2\right)^{-1}
\cr
&~~~~~~~~~~~~~~~~~~~~~~~~~~~~~~~~~~~~~~~~~~~~~~~~~~~~~~~~~~~~~
{}~~~~~~~~~~~~~~~~~~~~~\cdot (ln~g)^3 \biggr]^{-{g\over \lambda}}
\cr}
\eqno(28)$$
As $1\gg {1\over \lambda} > 0$, the second term in the bound (28) is a rapidly
decreasing function of the genus for large $g$, while the exponential
dependence of the first term on the genus is determined by $\epsilon_0^\prime$.
This parameter is constrained by modular invariance.  One of the conditions
defining the fundamental region of the modular group is $\vert det(C\tau +
D)\vert \ge
1$
for $\left(\matrix{A&B&
                    \cr
                    C&D&
                      \cr}\right) \in Sp(2g;{\Bbb Z})$.
Since
$$\vert det(C \tau+D)\vert~=~\vert det(Im~\tau)\vert~\vert det(C-iC(Re~\tau)
(Im~\tau)^{-1}-iD(Im~\tau)^{-1}\vert
\eqno(29)$$
the determinant will be greater than one for all C, D when $\vert
 det(Im~\tau)\vert~\ge~b~>~ 1$ for some number $b$,
so that $\vert det(C-iC(Re~\tau)(Im~\tau)^{-1}
-iD(Im~\tau)^{-1}) \vert$ will be bounded below when $det~C \ne 0$ and equal
to $\vert det~D\vert~\vert det(Im~\tau)^{-1}\vert~\ge~
\vert det(Im~\tau)\vert^{-1}$ when C=0.  Moreover, the minimum value of
${{tr(Im~\tau)}
\over g}$ is
$$\eqalign{ln~{1\over {\epsilon_0^\prime}}& -{1\over g}~\sum_{n=1}^g~s_{nn}
\cr
s_{nn}~is~a~least~upper&~bound~for~\sum_\alpha~^{(n,n)}  ~ln~\left\vert
{{\xi_{1n}-V_\alpha \xi_{2n}}\over {\xi_{1n}-V_\alpha \xi_{1n}}}
{{\xi_{2n}-V_\alpha \xi_{1n}}\over {\xi_{2n}-V_\alpha \xi_{2n}}} \right\vert
\cr}
\eqno(30)$$
implying that the restriction to a single fundamental domain requires
$$ln~{1\over {\epsilon_0^\prime}}~\ge~b^{1\over g}~+~{1\over g}~\sum_{n=1}^g~
s_{nn}
\eqno(31)$$
Thus, $\left[\left(ln~{1\over {\epsilon_0^\prime}}\right)^{-4}~-~
\left(ln~{1\over {\epsilon_0}}\right)^{-4}\right]~<~1$ and the first term in
the
bound (28) decreases exponentially.
{}From this result, it follows that the sum over genus of the moduli space
integrals defined by the class of Riemann surfaces associated with the
first three categories of isometric circles is finite.  Finiteness of
these integrals at large orders in perturbation theory contrasts with the
rapid divergence of the regularized vacuum amplitude in closed bosonic string
theory [2].

Another class of surfaces that might be included in the superstring path
integral are spheres with an infinite number of handles decreasing in size and
accumulating to a point.  These surfaces may be constructed using infinitely
generated groups of Schottky type, joining together isometric circles by
projective transformations in the extended complex plane [20].  Labelling the
handles by the index n, the decrease in the square of the radii of the
isometric circles, $\vert \gamma_n\vert^{-2} \sim {1\over
{n^{q^{\prime\prime}}}}$
characterizes the surfaces to be included in the path integral.
{}From the leading behaviour of the multiplier integral, the contribution of a
configuration of isometric circles with $q^{\prime\prime}_{\tilde N}$ given
in equation (9) may be estimated.
$$\eqalign{ \int_{\vert K_n\vert\sim {1\over {n^{q^{\prime\prime}_{\tilde
N}}}}}~
{{d \vert K_n\vert}\over { \vert K_n\vert \left(ln~{1\over {\vert
K_n\vert}}\right)^5}}
{}~&=~{1\over 4}[q^{\prime\prime}_{\tilde N}~ln~n~-~ln~\epsilon_0^\prime]^{-4}
{}~-~{1\over 4}[q^{\prime\prime}_{\tilde N}~ln~n~-~ln~\epsilon_0]^{-4}
\cr
{}~&=~(q^{\prime\prime}_{\tilde
 N} ~ln~n)^{-5}~ln~\left({{\epsilon_0^\prime}\over {\epsilon_0}}\right)
\cr
{}~&=~\left({{ln~n}\over {ln~\left({{\epsilon_0^\prime}\over
{\epsilon_0}}\right)}}
{}~+~{\tilde N}\right)^{-5} ~\left[ln~\left({{\epsilon_0^\prime}\over
{\epsilon_0}}\right)\right]^{-4}
\cr}
\eqno(32)$$
While the measure (4) indicates that the product over n of the integrals (32)
be evaluated, it is more useful to first sum over the index $\tilde N$ as this
will allow for the inclusion of all possible combinations of limits for
the multipliers and fixed points.

Having specified the limits for $\vert K_n\vert$, one may note that the
possible
values of $q^{\prime\prime\prime}$ consistent with non-overlap of the intervals
for $\vert \xi_{1n}-\xi_{2n}\vert$ are
$$q^{\prime\prime\prime}_{\tilde N^\prime}~=~{\tilde N^\prime}
{{ln \left({{\delta_0^\prime}\over {\delta_0}}\right)}\over {ln~n}}
{}~~~~~~~~~~~~{\tilde N^\prime}=0,~1,~2, ...~
\eqno(33)$$
As the super-fixed-point integral
gives $\vert \xi_{1n}-\xi_{2n}\vert^{-2}$, it will grow as $n^{2
q^{\prime\prime
\prime}}$, which will be shown to give rise to a divergence when
$q^{\prime\prime\prime}~>~{1\over 2}$.  However, it is also essential to
eliminate the overcounting of surfaces, which would arise when including every
value of $q^{\prime\prime\prime}_{\tilde N^\prime}$, $\tilde N^\prime$ integer.
In particular, surfaces corresponding to the limits ${{\delta_0}\over
{n^{2q^{\prime\prime\prime}}}}~<~\vert\xi_{1n}-\xi_{2n}\vert~<~
{{\delta_0^\prime}\over {n^{2q^{\prime\prime\prime}}}}$ can be obtained from
surfaces corresponding to ${{\delta_0}\over {n^{q^{\prime\prime\prime}}}}~<~
\vert \xi_{1n}-\xi_{2n}\vert~<~ {{\delta_0^\prime}\over
{n^{q^{\prime\prime\prime}}}}$ by pinching all handles other than those with
the index $n^2$ in the sequential labelling, removing the nodes and deforming
the surface by flattening the remaining portion of the handle on the sphere.
Pinching one handle produces a surface at the boundary of moduli space, or
equivalently, at the boundary of a string vertex, denoted by $\partial
{\cal M}_{g, 0}$ and $\partial {\cal V}_{g, 0}$ respectively at genus g [21].
Removing the two new punctures in the manner using an analytic map
[22] transforms the string vertex from ${\cal V}_{g-1,2}$ to ${\cal V}_{g-1,
0}$, which lies in the compactified moduli space ${\bar {\cal M}}_{g-1, 0}$.
At infinite genus, this procedure still produces a surface with an infinite
number of handles lying in ${\bar {\cal M}}_\infty$ [the number of ends of an
infinite-genus surface is not required here].
Thus the initial integration over the domain in ${\bar {\cal M}}_\infty$
associated with the configurations of isometric circles with the
range for $\vert \xi_{1n}-\xi_{2n}\vert$ given by $\left[{{\delta_0}\over
{n^{q^{\prime \prime\prime}}}},~{{\delta_0^\prime}\over
{n^{q^{\prime\prime\prime}}}}\right]$,  includes integration over the range
$\left[{{\delta_0}\over {n^{2q^{\prime\prime\prime}}}},~{{\delta_0^\prime}\over
{n^{2q^{\prime\prime\prime}}}}\right]$ up to pinching of handles.  The process
of pinching the handles is described, however, by the degeneration limits
$\vert K_m\vert \to 0,~m \ne n^2, n \in Z$.   It has already been shown that
these limits do not lead to any new divergences in the moduli space integral.
Integration over the range $\{\vert \xi_{1n}-\xi_{2n}\vert \sim {1\over
{n^{2q^{\prime\prime\prime}}}}\}$, therefore, only represents a finite addition
to  integration over the range $\{\vert \xi_{1n}-\xi_{2n}\vert \sim
{1\over {n^{q^{\prime\prime\prime}}}}\}$.

Since this argument can be repeated an arbitrary number of times, overcounting
will be eliminated by restricting the fixed-point integral to the range
$\{\vert \xi_{1n}-\xi_{2n}\vert \sim {1\over {n^{q^{\prime\prime\prime}}}}\},~
x>q^{\prime\prime\prime} > 0$, with x being a small positive number.  It
is therefore necessary to choose
$${{ln(ln~n)}\over {ln~\left({{\delta_0^\prime}\over {\delta_0}}\right)}}
{}~<~{\tilde N^\prime}_{max} \le x~{{ln~n}\over
{ln~\left({{\delta_0^\prime}\over
{\delta_0}}\right)}}
\eqno(34)$$
An upper bound for the integrals over the fixed points would then grow as
$n^{2x}$.

The constraints defining the fundamental region of the modular group,
$(Im~\tau)_{ss} \ge (Im~\tau)_{rr},~s\ge r$ generally reduce the integration
range of the Schottky group variables.  However, they are only relevant
when the intervals $\left[{{\epsilon_0}\over {n_s^{q^{\prime\prime}}}},~
{{\epsilon_0^\prime}\over {n_s^{q^{\prime\prime}}}}\right]$ and
$\left[{{\epsilon_0}\over {n_r^{q^{\prime\prime}}}},~
{{\epsilon_0^\prime}\over {n_r^{q^{\prime\prime}}}}\right]$ overlap, implying
$${{n_s}\over {n_r}}~\le \left({{\epsilon_0^\prime}\over
{\epsilon_0}}\right)^{1\over {q^{\prime\prime}}}
\eqno(35)$$
Assuming temporarily that the genus is finite and an integer between
$\left({{\epsilon_0^\prime}\over
 {\epsilon_0}}\right)^{{y-1}\over {q^{\prime\prime}}}$ and
$\left({{\epsilon_0^\prime}\over {\epsilon_0}}\right)^{y\over
{q^{\prime\prime}}}$, for some integer $y$,
the restrictions on each group of multipliers $K_n$ with $n$ lying between the
numbers
$1,~ \left({{\epsilon_0^\prime}\over {\epsilon_0}}\right)^{1\over
 {q^{\prime\prime}}},~\left({{\epsilon_0^\prime}\over {\epsilon_0}}\right)^{2
\over {q^{\prime\prime}}},~...$ leads to a combinatorial factor
$$\eqalign{&{1\over {\left(\left({{\epsilon_0^\prime}\over
{\epsilon_0}}\right)^{1\over
{q^{\prime\prime}}} -1\right)!~\left(\left({{\epsilon_0^\prime}\over
{\epsilon_0}}\right)^{2\over
{q^{\prime\prime}}}-\left({{\epsilon_0^\prime}\over
{\epsilon_0}}\right)^{1\over {q^{\prime\prime}}}\right)!~ ...~\left(g-
\left({{\epsilon_0^\prime}\over {\epsilon_0}}\right)^{{y-1}\over
 {q^{\prime\prime}}}\right)!}}
\cr
&~~~~~~~~~~~~~~~=~{{g! \left( \left({{\epsilon_0^\prime}\over {\epsilon_0}}
\right)^{y\over {q^{\prime\prime}}} -\left({{\epsilon_0^\prime}\over
{\epsilon_0}}\right)^{{y-1}\over {q^{\prime\prime}}}\right)!}
\over {\left({{\epsilon_0^\prime}\over {\epsilon_0}}\right)^{y
\over {q^{\prime\prime}}} ! \left(g- \left({{\epsilon_0^\prime}\over
 {\epsilon_0}}\right)^{{y-1}\over {q^{\prime\prime}}}\right) !}}
\left({{\left({{\epsilon_0^\prime}\over {\epsilon_0}}\right)^{1\over
{q^{\prime\prime}}}}\over {\left(\left({{\epsilon_0^\prime}
\over {\epsilon_0}}\right)^{1\over {q^{\prime\prime}}}
-1\right)^{1-\left({{\epsilon_0}\over {\epsilon_0^\prime}}\right)^{1\over
 {q^{\prime\prime}}}}}}\right)^{{\left({{\epsilon_0^\prime}
\over {\epsilon_0}}\right)^{{y+1}\over {q^{\prime\prime}}} -1}
\over {\left({{\epsilon_0^\prime}\over {\epsilon_0}}\right)^{1\over
{q^{\prime\prime}}}-1}}
{}~{1\over {g!}}
\cr
&~~~~~~~~~~~~~~~<~{1\over {g!}}~2^{{\left({{\epsilon_0^\prime}\over
{\epsilon_0}}
\right)^{2\over {q^{\prime\prime}}} g ~-~1}\over
{\left({{\epsilon_0^\prime}\over {\epsilon_0}}\right)^{1\over
{q^{\prime\prime}}} g - 1}}
\cr}
\eqno(36)$$
Defining the variable
$$t_{\tilde N}~=~\left({{\epsilon_0^\prime}\over {\epsilon_0}}\right)^{1\over
{q^{\prime\prime}_{\tilde N}}}~\simeq~\left({{\epsilon_0^\prime}\over
{\epsilon_0}}\right)^{{1\over {\tilde N}} {{ln~n}\over {ln~{{\epsilon_0^\prime}
\over {\epsilon_0}}}}}~=~n^{1\over {\tilde N}}
\eqno(37)$$
one finds that the combinatorial factor (37) is less than ${1\over
 {(g!)^{t_{\tilde N}-1}}}$.

The sum of the fixed-point integrals over all values of ${\tilde N^\prime}$
up to the maximum value in equation (34) grows as
$$\left({1\over {\delta_0^2}}-{1\over {\delta_0^{\prime 2}}}\right)
{}~\sum_{\tilde N^\prime =0}^{{\tilde
 N^\prime}_{max}}~n^{2q^{\prime\prime\prime}_{\tilde N^\prime}}~\simeq~{1\over
{\delta_0^2}}~n^{2x}
\eqno(38)$$
Multiplying this sum with the estimate (32) and dividing by the combinatorial
factor arising from the action of the modular group gives
$$\eqalign{{1\over {(g!)^{t_{{\tilde
 N}^{\prime}_{max}}-1}}}&~\left(ln~{{\epsilon_0^\prime}\over
{\epsilon_0}}\right)^{-4g}
{1\over {\delta_0^{2g}}} ~\prod_{n=1}^g~n^{2x}~\zeta \left(5, {{ln~n}\over
{ln~{{\epsilon_0^\prime}\over {\epsilon_0}}}}-1\right)
\cr
&\to ~ {1\over {(4 \delta_0^2)^g}} \prod_{n=1}^g~{1\over {(ln~n)^4}}
\cr}
\eqno(39)$$
in the limit $g\to \infty$ when $n^{1\over {{\tilde N}^\prime_{max}}}-1=2x$.
Since it is assumed that $x < 1$, the upper limit in equation (34) implies
that the last expression in equation (39) will be an adequate bound if
$x \le {1\over {\sqrt 2}} \left(ln \left({{\delta_0^\prime}\over {\delta_0}}
\right)\right)^{1\over 2}$.  As the product in (39) tends to zero as $g \to
\infty$, the integrals corresponding to the range
$\vert \xi_{1n}-\xi_{2n}\vert \sim {1\over {n^{2q^{\prime\prime\prime}_{\tilde
N^\prime}}}},~q^{\prime\prime\prime}_{\tilde N^\prime} > {1\over {\sqrt 2}}
\left(ln \left({{\delta^\prime_0}\over {\delta_0}}\right)\right)^{1\over 2}$
would be the limit of an exponential function of the genus for each value of
${\tilde N^\prime}$.
Thus, the magnitude of that part of the moduli space integral including spheres
having $g$ handles, with a decreasing cross-sectional area in the intrinsic
metric, near {\it a single} accumulation point, is therefore a countable sum
of contributions depending exponentially on the genus as $g\to \infty$.

For $q^{\prime\prime\prime}_{\tilde N^\prime} > {1\over {\sqrt 2}}
\left(ln \left({{\delta_0^\prime}\over {\delta_0}}\right)\right)^{1\over 2}$,
and particularly for $q^{\prime\prime\prime}_{\tilde N^\prime} > {1\over 2}$,
one initially obtains a divergence such as
$(g!)^{2q^{\prime\prime\prime}_{\tilde N^\prime}-1}$ in the integral, which
would be expected when confining attention to the Neveu-Schwarz sector of the
superstring.  However, this factorial divergence
has been avoided in the previous analysis because the behaviour of the
integrals is determined by the $\vert K_n\vert \to 0$ degeneration limit.
As noted earlier, this leads to an exponential dependence on the genus when
the GSO projection is used, or when the entire superstring spectrum, including
the Ramond sector, is included.

These conclusions are consistent with an analysis of the integrals of the
multipliers $B_m$ and $H_m$.  Like the $K_n$ integrals, a divergence of
factorial type would arise using the measure (23) appropriate for the
Neveu-Schwarz sector.  However, upon use of the full superstring measure,
this divergence must disappear, confirming the finiteness properties of
superstring amplitudes.  It may be recalled that the finiteness of genus-g
superstring amplitudes has been demonstrated by sewing together three-punctured
spheres, labelling the radii $R_j,~j=1,...,3g-3+N$ of the sewed punctures as
$K_n$, $B_m$, $H_m$ and $L_i$ (for external legs), and considering the limit
as $\vert R_j\vert \to 0$ [6].   For the limits $\vert B_m\vert,~\vert
H_m\vert \to 0$, a
finite result is obtained after replacing picture-changing operators by
commutators of BRST charges with bosonized ghosts and pulling the BRST charge
through the Beltrami differential for $K_n$, leaving a finite integral at the
boundary $\vert K_n\vert=0$ and an amplitude, involving picture-changing
operators
at a single puncture, which must vanish by the cancellation of fermion zero
modes [6].  By a parallel argument, similar to the one given before
equation (34), it follows that a finite, bounded integral is obtained in the
$B_m$ or $H_m$ degeneration limit for one connecting tube, and therefore an
exponential dependence on the genus should occur upon evaluation of all of the
integrals containing $\{B_m\}$ and $\{H_m\}$.
Similarly, the inclusion of the contribution of special surfaces, with handles
of arbitrarily small thickness, to
the {\it superstring} path integral, necessitated by the absence of a cut-off
on closed geodesic lengths, does not significantly alter the genus-dependence
of the bound on the amplitudes.

The bounds obtained above depended essentially on the restriction of the
integral over super-Schottky group parameter
space to a single fundamental domain of the super-modular group.  In this
connection, one may note that the super-mapping class group does not involve
any discrete transformations in the odd directions [23][24]. The relevant
constraints are those listed for the multipliers and ordinary fixed points.
Moreover, conditions such as $-{1\over 2}~\le~(Re~\tau)_{mn}~\le~{1\over 2}$
and $(Im~\tau)_{1n} \ge 0$ reduce the integrals by an exponential function of
the genus [15]. Other exponential factors arise from the angular integrations
of the arguments of $K_n$ and $\xi_{1n}-\xi_{2n}$, the integrations over
$\xi_{2n}$ and the primitive-element products in the measure (4).   Although
the powers in the primitive-element products in this measure differ from those
in the bosonic
string measure, the bounds obtained in [4] suggest that the products in
superstring measure will also be bounded by an exponential function of the
genus. Finally, a sphere with an infinite number of handles may have more
than one accumulation point, and generically, an infinite-genus Riemann surface
will have a Cantor set of ends [25], and, if these surfaces can be consistently
included in the superstring path integral, the counting of these surfaces would
affect the estimate (39) again by an exponential factor.

An alternative calculation switching the roles of the magnitudes of the
multipliers  $\vert K_n\vert$ and the distances between the fixed points $\vert
\xi_{1n}-\xi_{2n}\vert$ so that the values of ${\tilde N}$ are limited while
the values of ${\tilde N^\prime}$ are allowed to be arbitrarily large gives a
divergent result, but it is not related to the moduli space integral above,
because the elimination of handles on the surface, required in placing an upper
limit on ${\tilde N}$, involves the degeneration limit
$\vert \xi_{1m}-\xi_{2m}\vert \to 0$, with $m \ne n^2$ for example, rather than
$\vert K_m\vert \to 0$, as the magnitude of $\vert K_n\vert$ has already been
fixed by the choice of ${\tilde N}$.  Since the degeneration limit leads to
divergences, the pinching procedure relating configurations of isometric
circles of different ${\tilde N}$ does not involve a finite change to the
integrals.  Thus, the bounds with this range for ${\tilde N}$ and
${\tilde N^\prime}$ are not related to the estimates given in equation (39).

It may also be noted that the non-vanishing of
the integrals for finite $g$ is not inconsistent with the vanishing of the
superstring amplitudes with less than four massless external states.
The integral of the complete measure, including the Ramond sector, over the
entire integration range, would be required for a direct comparison.  Moreover,
it is sufficient to prove finiteness, removing the need to introduce a cut-off
near the boundary of moduli space, as the BRST contour integral arguments can
then be used to prove the non-renormalization theorems [26][27][28].
Similarly, even though the Euler character of moduli space grows asymptotically
as $(-1)^g {{(2g-1)!}\over {2^{2g-1} \pi^{2g}}}$ [29], the genus-dependence of
the volume is affected to a greater extent by the choice of measure.  While
there is a rapid, nearly factorial growth of the volume following integration
of the bosonic string measure over a subset of moduli space [4], the
genus-dependence of the corresponding integral is changed significantly by use
of the superstring measure (4).

These estimates do depend on the validity of applying the measure in
super-Schottky group coordinates to an integral over all of supermoduli space.
Integration over the odd moduli in superstring amplitudes is known to produce a
density over moduli space which possesses a total derivative ambiguity
associated with a change in the choice of the basis of super-Beltrami
differentials [30][31][32].  This integration ambiguity is related to the lack
of positive semi-definiteness of the superstring measure on a good global
slice, which follows from the absence of a global holomorphic section of the
universal Teichmuller curve [31].  Moreover, integration over the odd
moduli requires that there is a splitting of supermoduli space into
even and odd coordinates. The global obstruction to the splitting of 
supermoduli space can be circumvented by removing a divisor $D_g$ of 
codimension greater than or equal to one from the stable
compactification of moduli space, ${\bar {\cal M}}_g$.
Integration of a measure with the holomorphic factorization property [33][34]
is possible, when it is defined over a subset of moduli space and in particular
for the ranges of variables considered earlier.  The integral over all of
moduli space in superstring scattering amplitudes would differ in the
large-genus limit, therefore, from the estimates based on these subdomains by a
contribution from the divisor.  This, in turn, has been related to
tadpoles of massless physical states at lower genera [32].  Tadpole  
diagrams vanish in stable vacua and more generally, if an exponential 
dependence on the genus of the scattering amplitudes is assumed up to
genus $g$-1, the contribution from the boundary of moduli space should
also be bounded by a function of the same order, and it follows that
this dependence would continue to hold at genus $g$.
By induction, the exponential dependence should then be valid for arbitrarily
large genus.  Finally, there exist other formalisms which avoid the multi-loop
ambiguity although they are not directly related to the approach based on a
super-Schottky parametrization of supermoduli space.  In the light-cone
supersheet formalism, the boundary in supermoduli space is determined by the
requirement that the bosonic moduli are pure complex numbers without nilpotent
parts [35], eliminating the ambiguity arising from integration over
Grassmann variables.  Finite, unambiguous scattering amplitudes can also be
defined in the twistor-string formalism, which makes use of space-time
supersymmetry generators that are independent of the bosonized
super-reparametrization ghost fields having poles given by total derivatives in
moduli space [28], and it has been suggested previously that the exponential
dependence on the genus could be derived after considering the various
degeneration limits of these amplitudes.

The vanishing of N-point amplitudes for $N < 4$ provides the first indication
that the superstring amplitudes do not necessarily grow at a factorial rate
with respect to the genus.   Similarly, vanishing of superstring amplitudes
has been established when $g+N \le 8$ [36].  Until recently, however, it has
been difficult to determine the superstring scattering amplitudes at high
genus.
The only result that has been established thus far has been a consequence of
a technical argument relating N=2 string theories with topological field
theories.  This connection implies that there is a relation between special
type II string amplitudes in orbifold and Calabi-Yau backgrounds and
topological
string amplitudes at any given genus [37][38].
It is therefore of interest to be able to estimate the generic superstring
scattering amplitude with an arbitrary number of vertex operators, receiving
contributions from all genus.  This follows from the bounds in this paper
because the N-point g-loop scattering amplitude is typically given by
$$A_{N,g}~=~\int_{sM_g}~d\mu_g~\int_{s\Sigma_g}~dt_1 d{\bar t}_1 ...
dt_N d{\bar t}_N~ \langle V_1(t_1,{\bar t}_1) ... V_N(t_N, {\bar t}_N) \rangle
\eqno(40)$$
which may be bounded by
$$ \left\vert \int_{sM_g -{\cal N}(D_g)}~d\mu_g~\right\vert~
\left \vert~\int_{s\Sigma_g}~dt_1 d{\bar t}_1 ... dt_N d{\bar t}_N
{}~\langle V_1(t_1, {\bar t}_1) ... V_N(t_N, {\bar t}_N) \rangle ~\right\vert
\eqno(41)$$
where ${\cal N}(D_g)$ is a neighbourhood of the compactification divisor and a
subset of supermoduli space, which has measure zero in the large genus limit.
For the $R=-1$ slice of Teichmuller space,
$${1\over {2 \pi}}~\int_{\Sigma_g}~d^2 \xi~{\sqrt g}~R~=~2~-~2g
\eqno(42)$$
and using a subtraction procedure [39] to remove divergences
in the correlation function associated with the coincidence of vertex operators
and to bound its magnitude
\hfil\break
$ \vert \langle V_1 (t_1, {\bar t}_1) ... V_N (t_N,{\bar t}_N) \rangle \vert
{}~\le V_N^{max}$, it follows that the upper bound (41) is less than
$$(4 \pi (g-1))^N \cdot V_N^{max} \cdot B_K^g B_H^{g-2} B_B^{g-1}
\eqno(43)$$.  Lower
bounds of this type may also be derived provided there is a suitable lower
bound that can be used for expressions containing the supermoduli space
integral.  Since this particular integral, after the sum over spin structures,
actually vanishes, the lower bound that should not be determined strictly by
the
magnitude of this integral; rather, it should allow for the weighting of the
spin structures to be altered in the integration of the correlation function
over supermoduli space, leading to a non-vanishing amplitude.  It may be noted
that the bound (43) involved estimates of N-point functions on surfaces which
lie at the boundary of moduli space.  In the infinite-genus limit, one is led
to consider N-point functions on spheres with an infinite number of handles
of decreasing size.  One class of N-point functions has been studied in [20].

The genus-dependence of the special scattering amplitudes can be obtained
from the general estimates (43).  Specifically, the genus-g type II string
amplitude $A_{2g,g}$ in an orbifold or Calabi-Yau background, with
2g vertex operators including 2g-2 graviphotons and 2 gravitons, is equal to
$$A_{2g,g}~=~(g!)^2~F_g
\eqno(44)$$
where $F_g$ is the partition function of a topological string theory.
The partition function for a topological field theory defined
over a particular manifold M typically is
given by the product of the partition function over the submanifolds $M_i$,
such that M is the topological sum $\cup_i M_i$, so that the partition function
for a genus-g surface $Z_g$ would be $c^g Z_1^g$.  For a topological gravity
theory, the genus-g partition function will also involve a contribution from
a supermoduli space integral associated with modular deformations along the
collars joining the tori.  Since the magnitude of this supermoduli integral
has already been estimated, the genus-dependence of the partition function
$F_g$ should still be exponential, and indeed, this has been established in
several papers [38][40][41].  Consequently, the type II string amplitude is
$A_{2g,g}~=~c_1c_2^g (g!)^2$, consistent with the bound (44) upon
setting N equal to 2g.

The results obtained here imply, therefore,  there is a higher degree
of finiteness in the superstring theory in the large-genus limit when modular
invariance is considered.  The upper bound (28) decreases
exponentially with respect to the genus,
indicating that there may exist an exponential
bound for the higher-point superstring amplitudes.  In this case, the
perturbation series could be made to converge for an appropriate choice of
the string coupling constant or dilaton expectation value.
It may be noted that the estimates in this analysis are based on an exhaustion
of moduli space in the large-genus limit.  A further contribution to the
moduli space integrals, with a different dependence on the genus, is
conceivable, but the corresponding ranges of the super-Schottky group
parameters at large genus have not been considered relevant in this
approach.  Furthermore, improved summability of the perturbation series would
be associated generally with stability of the superstring vacuum.  This
property might be verified independently
by applying positive-energy theorems to the corresponding supersymmetric
backgrounds, in particular ${\Bbb R}^{10}$.

The estimates given here can also be regarded as consistent with the
addition of the exponential non-perturbative effects associated with the
joining of boundaries to the worldsheets.
The amplitudes corresponding to surfaces with boundaries attached behave as
$exp(-{1\over {\kappa_{str}}})$[42].
When Dirichlet boundary conditions are applied to all of the coordinates, the
boundary can be mapped to a single point and corresponds to a D-instanton.  The
one-D-instanton amplitude is given by
$$A_1~=~exp(<1>_{D_2}~+~...)~ A_1^{conn.}
\eqno(45)$$
where $<1>_{D_2}$ is the disk amplitude with no vertex operators and has weight
$-{1\over {\kappa_{str}}}$ [42].  It may be noted that string divergences,
arising when the vertex operators approach the Dirichlet boundaries, are
eliminated by a
Fischler-Susskind mechanism, in which amplitudes, associated with different
numbers of boundaries cancel [42][43].
Exponential non-perturbative effects
of the order of $exp(-{1\over {\kappa^2}})$ typically arise in quantum field
theories as a result of non-Borel summability of the
perturbation series [44].  Here the non-perturbative amplitudes can be
regarded as a separate contribution to the sum over surfaces, to be added
to the amplitudes corresponding to the closed surfaces.  A complete formulation
of string theory requires a sum over specific boundary types, and the
leading non-perturbative effects result, for example, from the addition of
Dirichlet boundaries.  Distinct boundary types also occur in the separation of
categories of surfaces into $O_G$ and $P_G$ surfaces, studied in the
infinite-genus limit of the S-matrix expansion [4][20].  Further progress
on a formula for the S-matrix including the combinatorics of boundaries on
the string worldsheets has been recently been reported in [45].  The scattering
amplitudes defined in this article arise as the logarithm of the first
term in the series expansion of the generating functional
$$ln~{\cal Z}~=~S^{(0)}~+~ln~\left[~\sum_n~{1\over
{n!}}\left(\int\prod_{i=1}^n~d^D
y_i^\mu
\right)~e^{S^{(n)\prime}}(y_1,...,y_n)~\right]
\eqno(46)$$
where $S^{(0)}$ is the path integral over connected worldsheets with no
boundaries, $S^{(n)}$ is the functional integral over worldsheets with
boundaries fixed at n points $y_i$ and $S^{(n)\prime}$ is $S^{(n)}$ without
the zero-boundary term.

The exponential bounds for the amplitudes, based only on the sum over closed
surfaces, are therefore not inconsistent, in this scheme, with the
non-perturbative effects that have been expected to be present in string
theory.  Further support for the estimates of superstring amplitudes in this
article comes from  the convergent perturbation expansion for QED with fermions
and finite ultraviolet and infrared cut-offs [46].  From the work on bosonic
string theory, one sees that the ultraviolet cut-off in this theory
is equivalent to the Gross-Periwal cut-off, while the infrared cut-off
removes the divergences in the infrared limit, analogous to the effect of
supersymmetry on string amplitudes.  It is conceivable that this version of
the QED model may arise in the low-energy limit of superstring theory,
providing an explanation for the dependence of the amplitudes on the
loop order.

It has recently been shown using duality in supersymmetric gauge theories that
the strong-coupling regime defined by electric variables is related to the
weakly-coupled regime based on magnetic monopoles [47][48].  This approach
might also be relevant in developing a complete, non-perturbative formulation
of QED, free of divergences of the charge-charge coupling at small length
scales [49].  A step towards such a theory has been made by showing that
the coupling of an electric charge to lines of magnetic flux increases less
rapidly, as the length scale decreases, than the coupling of two electric
charges [49], thus reminiscent of the phase of supersymmetric gauge theories
corresponding to a weakly-interacting system of magnetic monopoles
and abelian photons.

These results also have an analogue in superstring theory [50][51][52].
Specifically, there is a transformation in the S-duality group of
the type II superstring and heterotic string theories interchanging electric
and magnetic fields, while mapping the
dilaton field $\Phi$ to -$\Phi$, so that the string coupling constant
$\kappa_{str}^2 = <e^{2\Phi}>$ is mapped to its inverse 
${1\over {\kappa_{str}^2}}
= <e^{-2\Phi}>$ [53].  This suggests it may be feasible to completely determine
the S-matrix for the type IIB superstring, since calculations of scattering
amplitudes can be
done explicitly in the weak coupling limit.  Recently, it has been shown also
that the strong-coupling limit of a ten-dimensional type IIA string theory is
related to the weak-coupling
limit of eleven-dimensional supergravity [54].  One might therefore be able to
improve on earlier studies of quantum amplitudes for the eleven-dimensional
theory.  Moreover, it is known that
certain soliton states of both type II superstring theory and heterotic
string theory, saturating the Bogomol'nyi bound and forming representations of
$SL(2;{\Bbb Z})$, may be identified with extreme black holes [53][55].  The
extreme black hole solutions, which possibly could be used to describe
stable elementary particles [55][56], also represent background geometries for
which the amplitudes could be evaluated.  Conclusions about the bounds for the
amplitudes may determine certain properties of the states associated with
extreme black holes.

Finiteness of superstring amplitudes at any given order of perturbation theory
implies that a cut-off does not need to be introduced in moduli space and
and that ambiguities associated with surface terms arising from total
derivatives resulting from changes in the locations of picture-changing
operators are eliminated [6][31][57].  In particular, vanishing of the vacuum
amplitude at each order in perturbation theory follows from its representation
as the integral of a total derivative on the compactified moduli space
${\bar {\cal M}_g}$.  Similarly, the full vacuum
amplitude, given by a sum of multi-loop amplitudes, may be regarded as the
integral of a total derivative on the universal moduli space [22][24]
${\bar {\cal R}} = \prod_{g=0}^\infty \left(\cup_{k=0}^\infty~Sym^k({\bar
{\cal M}}_g)\right)$, and thus, it will be
determined entirely by its behaviour at the boundaries.   It is
conceivable that sums over surfaces with Dirichlet boundaries and infinite genus
might provide a non-vanishing  but not divergent contribution to 
`vacuum' amplitude, signalling the breakdown of supersymmetry.  There might exist a mechanism then for
breaking supersymmetry without affecting finiteness of the superstring
amplitudes.  Moreover, depending on the structure of ${\bar {\cal M}}_\infty$,
the combination of finiteness and supersymmetry breaking may select a specific
value of the string coupling constant, which could then be compared with
phenomenological values arising in unification models.

The results in this paper represent a further demonstration of the finiteness
of superstring theory.  While previous studies have been concerned with
finiteness
at any given order,  it is shown here that the leading behaviour of particular
moduli space integrals appears to depend only exponentially on the genus.
These conclusions were derived
using the super-Schottky group parameters as coordinates on supermoduli space
and restricting the integration region to a single fundamental domain of the
super-modular group.  The physical explanation for the removal of the factorial
dependence on the genus in superstring amplitudes is based on a connection
between infrared and large-order divergences, following from the
genus-dependence of the limits for the Schottky group parameters and
leading to their simultaneous elimination as a consequence of supersymmetry.
The estimates in this paper have been derived with the
purpose of bounding superstring amplitudes and reflect the absence of the
tachyon in the superstring spectrum, even though the full superstring
measure is not used.  The exponential dependence of the moduli space integrals
here, evaluated for a
specific class of surfaces, suggests that a similar dependence on the genus
should occur in full superstring theory, and this is supported by a general
argument
concerning the degeneration limits of superstring amplitudes.
The effect of supersymmetry on large-order string perturbation theory is a
greater degree of finiteness of the amplitudes, leading to an improvement of
the
summability of the loop expansion for superstring interactions.

\vskip 1in
\centerline{\bf Acknowledgements.}
I would like to thank Prof. S. W. Hawking and Dr. G. W. Gibbons for their
support while this research has been undertaken.  Useful discussions with
Dr. N. Berkovits, J. Briginshaw, Prof. M. Green, Dr. C. M. Hull,
Prof. E. Martinec, Dr. K. S. Narain, Prof. A. Strominger and Dr. G. Thompson
on superstrings are gratefully acknowledged.

\vfill
\eject

\centerline{\bf REFERENCES}
\item{[1]}  M. Kaku, ${\underline {Strings,~Conformal~Fields,~and~Topology}}$
(New York: Springer-Verlag, 1991)
\item{[2]}  D. J. Gross and V. Periwal, Phys. Rev. Lett. ${\underline {60}}$
(1988) 2105 - 2108
\item{[3]}  S. Shenker, `The Strength of Non-Perturbative Effects in String
Theory,'
\hfil\break
 $\underline{Random~Surfaces~and~Quantum~Gravity}$, Proc. 1990
 Cargese Workshop, ed. by O. Alvarez, E. Marinari and P. Windey
(New York: Plenum Press, 1991) 191 - 200
\item{[4]}  S. Davis, Class. Quantum Grav. ${\underline {11}}$ (1994) 1185 -
1200
\item{[5]}  E. D'Hoker and D. H. Phong, Rev. Mod. Phys. ${\underline{61}}$
(1988) 917 - 1065
\item{[6]}  N. Berkovits, Nucl. Phys. ${\underline{B408}}$ (1993) 43 - 61
\item{[7]}  J. L. Petersen, J. R. Sidenius and A. K. Tollsten, Nucl. Phys.
${\underline {B317}}$ (1989) 109 - 146
\item{[8]}  B. E. W. Nilsson, A. K. Tollsten and A. Watterstam, Phys. Lett.
${\underline {B222}}$ (1989) 399 - 405
\item{[9]}  G. Cristofano, M. Fabbrichesi and K. Roland, Phys, Lett.
${\underline {B236}}$ (1990) 159 - 164
\item{[10]}  P. DiVecchia, K. Hornfeck, M. Frau, A. Lerda and S. Sciuto,
Phys. Lett. ${\underline {B211}}$ (1988) 301 - 307
\item{[11]}  J. L. Petersen, J. R. Sidenius and A. K. Tollsten, Phys. Lett.
${\underline {B213}}$ (1988) 30 - 36
\item{[12]}  A. Bellini, G. Cristofano, M. Fabbrichesi and K. Roland, Nucl.
Phys.
${\underline {B356}}$ (1991) 69 - 116
\item{[13]}  J. L. Petersen,
${\underline{Physics~and~Mathematics~of~Strings:}}$
\hfil\break
${\underline{Memorial~Volume~for~Vadim~Knizhnik}}$,
ed. by L. Brink, D. Friedan and
\hfil\break
 A. M. Polyakov (Singapore: World Scientific, 1990)
\item{[14]}  C. L. Siegel, ${\underline {Topics~in~Complex~Function~Theory}}$,
Vol. 3 (New York: Wiley, 1973)
\item{[15]}  S. Davis, J. Math. Phys. ${\underline{36}}$(2) (1995) 648 - 663
\item{[16]}  G. Cristofano, R. Musto, F. Nicodemi and R. Pettorino,

              Phys. Lett. ${\underline {B217}}$ (1989) 59 - 65
\item{[17]}  D. Lust and S. Theisen, ${\underline{Lectures~on~String~Theory}}$
(Berlin: Springer-Verlag, 1989)
\item{[18]}  N. Seiberg and E. Witten, Nucl. Phys. ${\underline{B276}}$ (1986)
               272
\item{[19]}  N. Berkovits, private communication
\item{[20]}  S. Davis, Class. Quantum Grav. ${\underline 6}$ (1989) 1791 - 1803
\hfil\break
S. Davis, Mod. Phys. Lett. A, ${\underline 9}$ (1994) 1299 - 1307
\item{[21]}  B. Zwiebach, Nucl. Phys. ${\underline {B390}}$ (1993) 33 - 152
\item{[22]}  D. Friedan and S. Shenker, Phys. Lett. ${\underline {B175}}$
(1986) 287 - 295
\hfil\break
D. Friedan and S. Shenker, Nucl. Phys. ${\underline {B281}}$ (1987) 509 - 545
\item{[23]}  L. Crane and J. Rabin, Commun. Math. Phys. ${\underline {113}}$
(1988) 601 - 623
\item{[24]}  J. D. Cohn, Nucl. Phys. ${\underline {B306}}$ (1988) 239 - 270
\item{[25]}  L. Sario and M. Nakai, ${\underline {Classification~Theory~
of~Riemann}}$
\hfil\break
${\underline {Surfaces}}$ (Berlin: Springer-Verlag, 1970)
\item{[26]}  E. Martinec, Phys. Lett. ${\underline {B171}}$ (1986) 189 - 194
\item{[27]}  R. Kallosh and A. Morosov, Phys. Lett. ${\underline {B207}}$
(1988) 164 - 168
\hfil\break
A. Restuccia and J. G. Taylor, Phys. Rep. ${\underline {174}}$ (1989) 283 - 407
\item{[28]}  N. Berkovits, Nucl. Phys. ${\underline {B395}}$ (1993) 77 - 118
\item{[29]}  S. B. Giddings, E. Martinec and E. Witten, Phys. Lett.
${\underline {B176}}$ (1986) 362 - 368
\item{[30]}  J. J. Atick, J. M. Rabin and A. Sen, Nucl. Phys. ${\underline
B299}$ (1988) 279 - 294
\item{[31]}  J. J. Atick, G. Moore and A. Sen, Nucl. Phys. ${\underline
{B308}}$
(1988) 1 - 101
\item{[32]}  J. J. Atick, G. Moore and A. Sen, Nucl. Phys. ${\underline
{B307}}$
(1988) 221 - 273
\item{[33]}  P. Nelson, Commun. Math. Phys. ${\underline {115}}$ (1988) 167 -
175
\item{[34]}  G. S. Danilov, Class. Quantum Grav. ${\underline {11}}$ (1994)
2155 - 2161
\item{[35]}  N. Berkovits, Nucl. Phys. ${\underline {B304}}$ (1988) 537 - 556
\item{[36]}  O. Lechtenfeld and A. Parkes, Nucl. Phys. ${\underline{B332}}$
(1990) 39
\item{[37]}  I. Antoniadis, E. Gava, K. S. Narain and T. R. Taylor,
Nucl. Phys. ${\underline{B413}}$ (1994) 162 - 184
\item{[38]}  M. Bershdasky, S. Cecotti, H. Ooguri and C. Vafa, Commun. Math.
Phys. ${\underline{165}}$ (1994) 311 - 427
\item{[39]}  S. Hamidi and C. Vafa, Nucl. Phys. ${\underline{B279}}$ (1987)
465
\item{[40]}  H. Ooguri and C. Vafa, Harvard University preprint, HUTP-95/A017,
hep-th/9505183
\item{[41]}  N. Berkovits and C. Vafa, Nucl. Phys. ${\underline{B433}}$
(1995) 123
\item{[42]}  J. Polchinski, Santa Barbara preprint, NSF-ITP-94-73 (1994)
\item{[43]}  M. Gutperle, Cambridge preprint, DAMTP-95-07 (1995)
\item{[44]}  P. Ginsparg and J. Zinn-Justin, `Action principle and the large-
order behavior of non-perturbative gravity'
$\underline{Random~Surfaces~and~Quantum~Gravity}$,
\hfil\break
 Proc. 1990 Cargese Workshop
ed. by O. Alvarez, E. Marinari and P. Windey (New York: Plenum Press, 1991)
85 - 109
\hfil\break
P. Di Francesco, P. Ginsparg and J. Zinn Justin, Phys. Rep.
$\underline{254}$, 1 - 133
\item{[45]}  M. B. Green, CERN preprint, CERN-TH/95-78, hep-th/9504108
\item{[46]}  E. Caianello, Nuovo Cimento, ${\underline {3}}$ (1956) 223 - 225
\item{[47]}  N. Seiberg and E. Witten, Nucl. Phys. ${\underline {B426}}$
(1994) 19 - 52
\hfil\break
N. Seiberg and E. Witten, Nucl. Phys. ${\underline{B431}}$ (1994) 484 - 550
\item{[48]}  C. Vafa and E. Witten, Nucl. Phys. ${\underline{B431}}$ (1994)
3 - 77
\item{[49]}  A. S. Goldhaber, H. Li, R. R. Parwani, Stony Brook preprint,
ITP-SB-92-40 (1994)
\item{[50]}  M. J. Duff, Nucl. Phys. ${\underline{B335}}$ (1990) 610 - 620
\item{[51]}  J. H. Schwarz and A. Sen, Nucl. Phys. ${\underline{B411}}$
(1994) 35 - 63
\hfil\break
J. H. Schwarz and A. Sen, Phys. Lett. ${\underline{312B}}$ (1993) 105 - 114
\item{[52]}  A. Sen, Int. J. Mod. Phys. ${\underline{A9}}$ (1994) 3703 - 3750
\item{[53]}  C. M. Hull and P. K. Townsend, Nucl. Phys. ${\underline{B438}}$
(1995) 109 - 137
\item{[54]}  E. Witten, IAS preprint, IASSNS-HEP-95-18, hep-th/9503124
\item{[55]}  M. J. Duff and J. Rahmfeld, Phys. Lett. ${\underline{B345}}$
(1995)
441 - 447
\item{[56]}  M. J. Duff, R. R. Khuri, R. Minasian and J. Rahmfeld,
Nucl. Phys. ${\underline{B418}}$ (1994) 195 - 205
\item{[57]}  S. Mandelstam, Phys. Lett. ${\underline{B277}}$ (1992) 82 - 88

\end